\documentclass [prd,nofootinbib,eqsecnum]  {revtex4}
\usepackage{amssymb}
\usepackage{amsmath}
\usepackage{bm}
\usepackage{graphicx}

\newcommand{\hf}{{\hat{\phi}}}
\newcommand{\tf}{{\tilde{\phi}}}
\newcommand{\hx}{\hat{x}}

\newcommand{\hP}{\hat{\partial}}

\def\SO{\sf{SO}}
\def\he{{\hat e}}
\newcommand{\be}{\begin{equation}}
\newcommand{\ee}{\end{equation}}
\newcommand{\bea}{\begin{eqnarray}}
\newcommand{\eea}{\end{eqnarray}}

\begin{document}

\title{%
Lorentz invariant field theory on  $\kappa$--Minkowski space }
\author{Michele Arzano}
\email{marzano@perimeterinstitute.ca} \affiliation{Perimeter
Institute, 31 Caroline Street North Waterloo, Ontario, Canada N2L
2Y5}
\author{Jerzy Kowalski-Glikman}
\email{jkowalskiglikman@ift.uni.wroc.pl} \affiliation{Institute for
Theoretical Physics, University of Wroc\l{}aw, Pl.\ Maxa Borna 9,
Pl--50-204 Wroc\l{}aw, Poland}
\author{Adrian Walkus}
\email{walkus@ift.uni.wroc.pl} \affiliation{Institute for Theoretical Physics,
University of Wroc\l{}aw, Pl.\ Maxa Borna 9, Pl--50-204 Wroc\l{}aw, Poland}

\date{\today}
\begin{abstract}
It is by now well established that the momentum space dual to the non-commutative $\kappa$-Minkowski space is a submanifold of de Sitter space. It has been noticed recently that  field theories built on such momentum space suffer from a subtle form of Lorentz symmetry breaking. Namely, for any negative energy mode the allowed range of rapidities is bounded above.  In this paper we construct a complex scalar field theory with a modified action of Lorentz generators which avoids this problem.  For such theory we derive conserved charges corresponding to translational and $U(1)$ symmetries. We also discuss in some details the inner product and Hilbert space structure of the $\kappa$-deformed complex quantum field.

\end{abstract}

\maketitle

\section{Introduction}
Field theories on non-commutative space-times \cite{Doplicher:1994tu} have been a subject of intensive studies for almost a decade now.  Such theories are believed to represent an effective flat limit of quantum gravity in which the local degrees of freedom of gravity are suppressed and the only remaining trace of gravity is the presence of an energy scale related to Planck energy.  This expectation is supported by exact calculations in $2+1$ gravity coupled to a scalar field \cite{Freidel:2005bb}, \cite{Freidel:2005me}, where it has been shown that indeed integrating out the (topological in this case) degrees of freedom of gravity results in a non-commutative field theory. It has been also showed that non-commutative space-times can arise in the context of string theory \cite{Seiberg:1999vs}. Recently it has been argued on general grounds \cite{Arzano:2007nx} that non-commutativity and Hopf algebra symmetries arise in quantum field theories as a result of introducing certain (quantum) gravity induced non-local effects.\\
In this paper we will concentrate on a particular type of non-commutative space-time, the so-called $\kappa$-Minkowski space \cite{Majid:1994cy}, \cite{Lukierski:1993wx}\footnote{This formula features $\kappa$ explicitly; in what follows, to simplify notation, we set $\kappa=1$, which means that both positions and momenta are dimensionless, expressed in Planck units (assuming that $\kappa$ is Planck mass). We will however reintroduce $\kappa$ in some key formulas so as to make them more transparent.}
\begin{equation}\label{1}
 [\hat x^0, \hat x^i] = \frac{i}\kappa\, \hat x^i.
\end{equation}
This space arises \cite{Majid:1994cy} as a dual of the momentum sector of the $\kappa$-Poincar\`{e} algebra (see below (\ref{2}), (\ref{3})) \cite{Lukierski:1991pn}, \cite{Lukierski:1992dt} and as such is a natural candidate for describing space-time in the context of Doubly Special Relativity
(see \cite{Amelino-Camelia:2000ge}, \cite{Amelino-Camelia:2000mn},
\cite{KowalskiGlikman:2001gp}, \cite{Bruno:2001mw} for original formulation of DSR and
\cite{Kowalski-Glikman:2004qa}, \cite{Kowalski-Glikman:2006vx} for reviews.)\\
According to \cite{Majid:1994cy}  Lorentz symmetry acts on the noncommutative positions $\hat x^\mu$ in the standard way\footnote{In spite of the apparent non-covariance of the $\kappa$-Minkowski defining commutator (\ref{1}), it is in fact covariant, as a result of non-trivial co-product of boosts. See
\cite{Freidel:2007yu} for details.}
$$
 M_{i}\triangleright \hat{x}_{0}=0,\quad M_{i}\triangleright \hat{x}_{j}=i\epsilon_{ijk}\hat{x}_{k},\quad
 N_{i}\triangleright \hat{x}_{0}=i\hat{x}_{i},\quad N_{i}\triangleright \hx_{j}=i\delta_{ij}\hx_{0}
$$
and this action can be extended by duality to the action on momentum space.  As we will see this space is a group manifold and in a particular parametrization of (coordinates on)\footnote{These coordinates are usually called the `bicrossproduct basis'.} this space, the four ``momentum'' variables $k_\mu=(k_0, k_i)$ transform in a deformed (with deformation parameter $\kappa$), non-linear fashion, as follows
\begin{equation}\label{2}
  [M_i, k_j] = i\, \epsilon_{ijk} k_k, \quad [M_i, k_0] =0
\end{equation}
\begin{equation}\label{3}
   \left[N_{i}, {k}_{j}\right] = i\,  \delta_{ij}
 \left( \frac12 \left(
 1 -e^{-2{k_0}}
\right) + \frac{\mathbf{k}^2}2  \right) - i\, k_{i}k_{j} ,
\quad
  \left[N_{i},k_0\right] = i\, k_{i}
\end{equation}
Equations (\ref{2}), (\ref{3}) define the algebraic part of $\kappa$-deformed  Poincar\`e algebra, which becomes a Hopf algebra after  appropriate  structures are added (see below). This Hopf algebra is called  $\kappa$-Poincar\`e algebra and was first derived in \cite{Lukierski:1991pn}, \cite{Lukierski:1992dt}.\\
It has been known for some time that both $\kappa$-Minkowski space and $\kappa$-Poincar\`e algebra can be understood geometrically as resulting from the fact that the momentum space is curved, forming (a submanifold of -- see below) the four dimensional de Sitter space of curvature $\kappa$ \cite{KowalskiGlikman:2002ft},
\cite{Freidel:2007hk}. This space is defined as a four-dimensional hypersurface
\begin{equation}\label{4}
   -P_0^2 + P_1^2 + P_2^2 + P_3^2 + P_4^4 =1
\end{equation}
in five dimensional Minkowski (momentum) space. The bicrossproduct coordinates $k_\mu$ on this space are related to the global coordinates $P_A$ in (\ref{4}) as follows
\begin{eqnarray}
 {P_0}(k_0, \mathbf{k}) &=&  \sinh
{{k_0}} + \frac{\mathbf{k}^2}{2}\,
e^{  {k_0}}, \nonumber\\
 P_i(k_0, \mathbf{k}) &=&   k_i \, e^{  {k_0}}, \nonumber\\
 {P_4}(k_0, \mathbf{k}) &=&  \cosh
{{k_0}} - \frac{\mathbf{k}^2}{2}\, e^{  {k_0}}.
\label{5}
\end{eqnarray}
and one obtains Lorentz transformations (\ref{2}), (\ref{3}) just by noticing that in (\ref{4}) $P_\mu$, being Lorentzian coordinates on five dimensional Minkowski space, are components of a linearly transforming Lorentz vector, while $P_4$ is a Lorentz scalar. Here by Lorentz symmetry we mean the $\SO(3,1)$ subgroup of the full $\SO(4,1)$ symmetry group of (\ref{4}), with generators $M_i$, $N_i$ as above
\begin{equation}\label{6}
   \left[N_{i}, {P}_{j}\right] = i\,  \delta_{ij}\, P_0,
\quad
  \left[N_{i},P_0\right] = i\, P_{i}, \quad \left[N_{i},P_4\right] =0
\end{equation}
It follows from (\ref{4}) that $P_4$ is directly related to the standard Poincar\`e mass Casimir $P_4=\pm\sqrt{1+m^2}$.\\
One immediately sees that there is a clash between the fact that the momentum space is a submanifold of de Sitter space covered by $k$ coordinates and Lorentz symmetry, noticed already in the paper \cite{Bruno:2001mw}, and discussed in details in \cite{Freidel:2007hk} and \cite{Freidel:2007yu}. Namely, it follows  from (\ref{5}) that these coordinates cover just a half of the de Sitter momentum space (\ref{4}), defined by the inequality
\begin{equation}\label{7}
    P_+ \equiv P_0 + P_4 >0
\end{equation}
This condition is however manifestly not Lorentz-invariant: the orbits of Lorentz group action for negative energies ($P_0 <0$ or $k_0<0$) necessarily cross the boundary of this submanifold $P_+=0$. In other words, picking any negative energy state $P_0<0$, $|P_0|<P_4$ and acting on it with boosts, there always exist the value of the rapidity parameter for which $P_0$ becomes equal $-P_4$ and the inequality (\ref{7}) is violated.\\
One would think that this problem is just a coordinates artifact, but unfortunately this is not the case. The formulas for coproduct and antipode, the notions that are necessary tools in constructing field theory on $\kappa$-Minkowski space become singular at $P_+=0$ in all coordinates, which is not really surprising because it is the submanifold $P_+>0$ where the theory has been consistently defined in the first place. This condition characterizes therefore the part of de Sitter momentum space, on which field theory can be built.\\
To see this from a different perspective notice (see \cite{Freidel:2007hk} for details) that  an ``ordered plane wave on $\kappa$-Minkowski space'' \cite{Amelino-Camelia:1999pm}
\begin{equation}\label{8}
   \he_k \equiv e^{ik_i \hat x^i} e^{ik_0 \hat x^0}
\end{equation}
has a clear interpretation of being a Borel (also called sometimes $AN(3)$) group element, where the Borel group is defined by exponentiating the algebra (\ref{1}) regarded now as a Lie algebra\footnote{It is worth noticing that the metric on group manifold in coordinates (\ref{8}) is just the standard cosmological metric $-dk_0^2 + e^{2k_0}\, dk_i^2$.}.
In this parametrization the group structure is given by composition of non-commutative plane waves
\begin{equation}\label{9}
    \he_{kl} \equiv \he_k \he_l = e^{i\hat x^i(k_i + e^{-k_0}l_i)} e^{i \hat x^0(k_0 + l_0)}
\end{equation}
which, as one can check directly, preserves the condition $P_+>0$: $P_+(k_0+l_0, k_i + e^{-k_0}l_i)=e^{k_0+l_0}>0$. Note in passing that this group structure can be interpreted naturally in terms of the coproduct of $\kappa$-Poincar\`e algebra. Thus the momentum space dual to the $\kappa$-Minkowski space is nothing but the Borel group manifold, which is a submanifold of de Sitter space, again defined by the condition $P_+>0$. As we discussed above this manifold is not invariant under the natural linear action of Lorentz group. It follows that a theory with such momentum space and linear Lorentz group action on it will necessarily lead to Lorentz symmetry breaking, although this breaking is of a rather subtle, global nature and cannot be seen at the level of infinitesimal transformations (see \cite{Freidel:2007yu}).\\
This is the problem that we would like to address in this paper. In the following Section we propose that in order to resolve it we have to modify the action of Lorentz transformations, and we  argue that this modification is, after all, quite natural. In Section III we construct a complex scalar field incorporating this modified Lorentz action, and in Section IV we calculate its translational Noether charges. Alongside with these charges we derive the conserved charge associated with the $U(1)$ internal symmetry of the complex field.  In Section V we discuss the relation of such charge with the invariant inner product for the solutions of the field equations and use it to construct the quantum one-particle Hilbert spaces.  The last section will be devoted to conclusions.

\section{Negative energy, antipode, and Lorentz invariance}
It follows from the discussion in the preceding section that in order to secure Lorentz invariance one must change the way Lorentz transformation acts on negative energy modes.  Indeed for positive energy $k_0>0$, which is equivalent to $P_0>0$, the condition $P_+>0$ is clearly identically satisfied for positive $P_4 = +\sqrt{1+m^2}$, because Lorentz transformations do not change sign of the zero component of a Lorentz vector, and $P_0$ is always positive. For negative energies $k_0<0$ ($P_0<0$) however the condition $P_+>0$ becomes non-trivial and is not compatible with Lorentz invariance.\\
Let us start by considering a positive energy plane wave labeled by $k_\mu = (k_0, k_i)$, $k_0>0$
\begin{equation}\label{10}
    \he_k \equiv e^{ik_i \hat x^i} e^{ik_0 \hat x^0}
\end{equation}
It is natural to assume that its negative energy counterpart is described by a conjugated plane wave
\begin{equation}\label{11}
     (\he_k)^\dag = e^{-i  k_0 \hat{x}^0} e^{-i   k_i \hat{x}^i} = e^{-i   (e^{k_0}k_i)\hat{x}^i}e^{-i  k_0 \hat{x}^0} = \he_{S(k)}
\end{equation}
labeled by the generalized `minus' momentum, called in the language of Hopf algebras the antipode
\begin{equation}\label{12}
   S(k_i) = - e^{k_0}k_i, \quad S(k_0) = - k_0\, ,
\end{equation}
notice that $\he_k \, (\he_k)^\dag = 1$.\\
Let us now check how the antipode transforms under the action of Lorentz transformation. To this end it is convenient to use the parametrization in terms of $P$ variables. Since antipode is a homomorphism, one easily calculates
\begin{equation}\label{14}
    S(P)_0 \equiv P(S(k))_0 =-P_0+\frac{\mathbf{P}^2}{P_0+P_4}, \quad S(P)_i \equiv P(S(k))_i =-\frac{{P}_i}{P_0+P_4}, \quad S(P)_4 \equiv P(S(k))_4 = P_4
\end{equation}
Note in passing that the last equality along with $P_4 = +\sqrt{1+m^2}$ will ensure that in the quantum framework particle and antiparticle have the same mass. Using the identity
$$
S(P)_0 =-P_4+\frac{1}{P_0+P_4}
$$
One immediately sees that $S(P)_0 +S(P)_4$ is positive provided $P_+>0$.\\
To see this explicitly, let us take a negative energy state of mass $m$ that is initially at rest, with $P_0(0)=-S(P)_0(0) =m$. Using eq.\ (\ref{14}) and applying the boost with rapidity parameter $\xi$ we find that
\begin{equation}\label{15}
    S(P)_0(\xi) = \frac1{m\cosh\xi + \sqrt{1+m^2}} - \sqrt{1+m^2}\end{equation}
which is negative for any $\xi$, while
$$S(P)_0(\xi) + S(P)_4(\xi)=\frac1{m\cosh\xi + \sqrt{1+m^2}}$$
is positive and goes to zero for infinite boost.\\
The infinitesimal Lorentz transformations in $k$ parametrization  take the form
\begin{equation}\label{16}
   \left[N_{i}, S({k})_{j}\right] = i\,  \delta_{ij}
 \left( \sinh S(k)_0 + \frac{\mathbf{S}(k)^2}2\, e^{S(k)_0}  \right)  ,
\quad
  \left[N_{i},S(k)_0\right] = i\, S(k)_{i}\, e^{-S(k)_0}
\end{equation}
while in the $P$ parametrization one finds
\begin{equation}\label{16a}
   \left[N_{i}, S({P})_{j}\right] =- i\,  \delta_{ij}\left[P_4^2+P_4 +1\right] + S({P})_{i}S({P})_{j}
  ,
\quad
  \left[N_{i},S(P)_0\right] = i\, S(P)_{i}\, \left[ S(P)_0+P_4\right]\, .
\end{equation}
To summarize the results of this section: the momentum space is bounded for negative energies, and the bound is not Lorentz invariant if boosts act in the standard, linear way. However, we can parametrize the negative energy modes with the antipode and in this parametrization the action of Lorentz symmetry is nonlinear, preserving the bound. As we will see in the next section this parametrization is actually quite natural, consistent with properties of ordered plane waves on $\kappa$-Minkowski space.

\section{Plane waves and scalar field}
Field theory on $\kappa$-Minkowski space has been constructed and discussed in several papers including \cite{Kosinski:1999ix}, \cite{Kosinski:2001ii}, \cite{Amelino-Camelia:2001fd}, \cite{Kosinski:2003xx}, \cite{Daszkiewicz:2004xy}, \cite{Dimitrijevic:2003wv}, \cite{Dimitrijevic:2004nv}, \cite{Freidel:2006gc}. However it was only in the papers \cite{Freidel:2007hk}, \cite{Freidel:2007yu} where the problem of Lorentz symmetry breaking has been noticed and discussed. Here we base our discussion on the results of \cite{Freidel:2007hk}, which contains also the detailed construction and discussion of tools and techniques that we will be using below.\\
There are two basic tools which are required to construct a field theory on $\kappa$-Minkowski space: the space of basic functions -- plane waves and a differential calculus. In light of the results of the preceding section we propose to use different sets of plane waves for positive and negative energies
\begin{equation}\label{17}
    \he_k^+ \equiv e^{ik_i \hat x^i} e^{ik_0 \hat x^0}, \quad k_0 >0
\end{equation}
and
\begin{equation}\label{18}
    \he_k^- \equiv e^{-ik_0 \hat x^0}e^{-ik_i \hat x^i}= e^{i S(k)_i \hat x^i} e^{i S(k)_0 \hat x^0}, \quad  k_0 >0
\end{equation}
with $(\he_k^+)^\dag= \he_k^-$.\\
In terms of the functions $\he_k^\pm$ a scalar field will be given by
\begin{equation}\label{19}
    \hf(\hx) = \int_{k_0>0} d\mu(k) \left(\phi^+(k) \he_k^+(\hx) + \phi^-(k) \he_k^-(\hx)\right)
\end{equation}
where $d\mu(k)$ is a left invariant $d\mu(pk)=d\mu(k)$ and Lorentz
invariant measure on the Borel group (which is exactly a diffeomorphism invariant measure on de Sitter space in these coordinates -- see footnote 4 above)
\begin{equation}\label{19a}
d\mu(k) = \frac{e^{3k_0}}{(2\pi)^4}\, dk_0
d^3\mathbf{k}=\frac{1}{(2\pi)^4\,P_4}\, dP_0 d^3\mathbf{P}
\end{equation}
where the second equality can be straightforwardly derived by changing variables, see (\ref{5}). In the case of an on-shell field
$P_4=\sqrt{1+m^2}$ and thus the measure (\ref{19a}) differs from the standard Lorentz invariant measure only by a constant factor.
Notice that if the field $\hf(\hx)$ is real, i.e. $\hf(\hx)^\dag =\hf(\hx)$, one must have $\overline{\tf(k)^+}=\tf(k)^-$, where the bar denotes the complex conjugation, which is just the standard reality condition.\\
The next step is to introduce dynamics for the field defined above.  In order to do so we need a differential calculus which is compatible with the action of Lorentz symmetry. Details of construction of such calculus can be found in \cite{Freidel:2007hk} and references therein; here we will just recall the definition of objects that we will need in what follows. It turns out that the (bi-) covariant differential calculus associated with $\kappa$-Minkowski space structure (\ref{1}) is necessarily five-dimensional, i.e. it consists of five independent partial derivatives $\hP_A$, $A=0,\ldots,4$, whose action on a positive energy plane wave is given by
\begin{equation}\label{20}
   \hat \partial_\mu \he_k^+ = P_\mu\, \he_k^+, \quad \hat \partial_4 \he_k^+ = (1-P_4)\, \he_k^+
\end{equation}
where $P_A$ are  given by (\ref{5}), while
\begin{equation}\label{21}
   \hat \partial_\mu \he_k^- = S(P)_\mu\, \he_k^-, \quad \hat \partial_4 \he_k^- = (1-S(P)_4)\, \he_k^-=(1-P_4)\, \he_k^-
\end{equation}
where the antipodes are given by eq.\ (\ref{14}). Notice that these definitions feature no $i$ factor. The hermitian conjugate derivative is then defined by the identity
\begin{equation}\label{22}
    (\hP_A \he_k^+)^\dag\equiv \hP_A^\dag \he_k^-, \quad (\hP_A \he_k^-)^\dag\equiv \hP_A^\dag \he_k^+
\end{equation}
Having these definitions we can turn to the dynamics of our field.  We will consider a free massive scalar theory, given by the  Lagrangian
\begin{equation}\label{23}
   \hat {\cal L} =\left[ (\hP_\mu \hf)^\dag \hP^\mu \hf + m^2\hf^\dag \hf\right]\, .
\end{equation}
To derive the field equation following from (\ref{23}) consider the action
\begin{equation}\label{24}
    S=\int d^4\hat x \hat {\cal L}
\end{equation}
where the integral over $\kappa$-Minkowski space is defined in such a way that
$$
\int d^4\hat x e^{ik_i\hx_i}e^{ik_0\hx^0} \equiv (2\pi)^4 \delta^3(\vec{k})\, \delta(k_0)\, .
$$
Taking into account the decomposition (\ref{19}) along with the definitions (\ref{20})--(\ref{22}) and using the identities
$$
\int d^4\hat x \he^-_k\, \he^+_l= (2\pi)^4 \delta^3(-e^{k_0}\, \vec{k}+e^{k_0}\, \vec{l})\delta(l_0-k_0), \quad \int d^4\hat x \he^+_k\, \he^-_l= (2\pi)^4 \delta^3( \vec{k}-e^{l_0-k_0}\, \vec{l})\delta(k_0-l_0)
$$
and
$$
\int d^4\hat x \he^+_k\, \he^+_l=\int d^4\hat x \he^-_k\, \he^-_l=0\, ,
$$
where the last equality means that the integrals are proportional to $\delta(k_0+l_0)$ which, since the argument is strictly positive, is the zero distribution. We find
\begin{multline}
   S=\int d^4\hat x \hat {\cal L} = \int d\mu(k)d\mu(l)\left(P_\mu(k)P^\mu(l) + m^2\right)\bar\phi^+(k)\phi^+(l)\, \delta(e^{k_0}(\vec{l}-\vec{k}))\delta(l_0-k_0)\\
+\int d\mu(k)d\mu(l)\left(S(P(k))_\mu S(P(l))^\mu + m^2\right)\bar\phi^-(k)\phi^-(l)\, \delta(\vec{k}-\vec{l}))\delta(k_0-l_0)\, .
\end{multline}
Then, since $e^{k_0}$ is positive, and because $S(P)_\mu S(P)^\mu = P_\mu P^\mu$ the field equations for all modes are
\begin{equation}\label{26}
    P_\mu(k)P^\mu(k) + m^2=0
\end{equation}
Finally we can write the on-shell field, solution of the equation of motion, as
\begin{equation}\label{27}
    \hf(\hx)^{on-shell} = \int_{k_0>0} 2\pi\, d\mu(k)\,  \delta(P_\mu(k)P^\mu(k) + m^2)\, \left(\phi^+(k) \he_k^+(\hx) + \phi^-(k) \he_k^-(\hx)\right)\, .
\end{equation}
We will make use of this expression in the calculation of conserved charges in the following Section and in constructing the quantum field Hilbert space in Section V.

\section{Conserved charges}
For a classical field theory we have a powerful tool for characterizing symmetries.  Noether's theorem tells us that to each symmetry of the action we can associate a conserved charge and that in turn this charge provides a symmetry generator for the theory.  In the corresponding quantum theory such charges will be directly related to the observables. As stressed in \cite{Agostini:2006nc}, \cite{AmelinoCamelia:2007rn} conserved charges related to space-time symmetries are even more important in the context of deformed field theories, the reason being that in this latter case there does not exist any natural labeling of plane waves and therefore are the charges, and not the labels, that correspond to physically observed quantities.\\
In the discussion below we will follow the general results derived in \cite{Freidel:2007hk}. Since there are five independent derivatives $\hP_A$ there are also five independent translational charges
\begin{equation}\label{28}
    {\cal P}_A = - \int_{\mathbb{R}^3}\left(\hP_A\hat\Pi\hf +\hP_A\hf^\dag\hat\Pi^\dag\right)
\end{equation}
where the field is assumed to be on-shell and $\hat\Pi$ denotes the deformed ``conjugate momentum"
\begin{equation}\label{29}
    \hat\Pi\equiv(1-\hP_4)\hP_0\hf^\dag
\end{equation}
which differs from the time derivative of the field just by a constant multiplicative factor. Note that the expression for space time components $A=0,\ldots,3$ of the conserved charges looks formally exactly like in the case of the standard free scalar field on Minkowski space.\\
We postpone the explicit derivation of the form of the translational charges (\ref{28}) and instead we turn to another conserved object which we think is more interesting and useful. This is the conserved charge associated with the $U(1)$ global symmetry of our complex field.  Its importance stems from the fact that it is strictly connected with the natural symplectic form for classical solutions and thus it plays a vital role in defining the Hilbert space inner product in the quantum theory. This charge is given by
\begin{equation}\label{30}
    {\cal Q}=-\int_{\mathbb{R}^3}\left(\hat\Pi\hf +\hf^\dag\hat\Pi^\dag\right)\, .
\end{equation}
We want to write an explicit expression for ${\cal Q}$.  There will be contributions from products of plane waves of the type $\he_k^+\he_l^+$, $\he_k^+\he_l^-$ etc. Let us consider them term by term.  Integrating $\he_k^+\he_l^+$ over $\mathbb{R}^3$ we get
$$
\int_{\mathbb{R}^3} d^3x\,\he_k^+\he_l^+= \int_{\mathbb{R}^3} d^3x\,
e^{i(\vec{k}+e^{-k_0}\vec{l})\vec{x}}\, e^{i(k_0+l_0)x^0} = (2\pi)^3
\delta^3(\vec{k}+e^{-k_0}\vec{l})\, e^{i(k_0+l_0)x^0}
$$
Therefore this term contribution to the charge (\ref{30}) will involve the above delta along with the mass-shell conditions for $k$ and $l$ modes with the requirements that $k_0, l_0>0$. All together we have the set of equations
$$
P_0^2(k) - \vec{P}^2(k)=m^2, \quad P_0^2(l) - \vec{P}^2(l)=m^2
$$$$
\vec{l}=-e^{k_0}\vec{k}, \quad k_0 >0, \quad l_0 >0
$$
One can check that there are no solutions to this set of equations, and thus there is no contribution from this term to the expression (\ref{30}). Similarly there is no contribution coming from the integral of $\he_k^-\he_l^-$.\\
The integrals of ``mixed" products of plane waves $\he_k^+\he_l^-$, $\he_k^-\he_l^+$ are given by
$$
\int_{\mathbb{R}^3} d^3x\,\he_k^+\he_l^- = (2\pi)^3
\delta^3\left(\vec{k}-e^{-k_0+l_0}\vec{l}\right)\, e^{i(k_0-l_0)x^0},
\quad \int_{\mathbb{R}^3} d^3x\,\he_k^-\he_l^+ = (2\pi)^3
\delta^3\left(e^{k_0}(\vec{l}-\vec{k})\right)\, e^{i(l_0-k_0)x^0}\, .
$$
Terms involving such products will contribute to the charge which we write as ${\cal Q}={\cal Q}^{(1)}+{\cal Q}^{(2)}$ where
$$
{\cal Q}^{(1)}=-\int (2\pi)^5\, d\mu(k)
d\mu(l)\delta(P^2(k)+m^2)\delta(P^2(l)+m^2)P_4(k)\, \bar\phi^-(k)\phi^-(l)\,\delta^3\left(\vec{k}-e^{-k_0+l_0}\vec{l}\,\,\right)\,
e^{i(k_0-l_0)x^0}[P_0(k)+P_0(l)]\, .
$$
and
$$
{\cal Q}^{(2)}=-\int (2\pi)^5\, d\mu(k)
d\mu(l)\delta(P^2(k)+m^2)\delta(P^2(l)+m^2)P_4(k)\, \bar\phi^+(k)\phi^+(l)\,\delta^3\left(e^{k_0}(\vec{k}-\vec{l})\right)\,
e^{i(k_0-l_0)x^0}[S(P)_0(k)+S(P)_0(l)]\, .
$$
After a straightforward but rather tedious calculation one finds
$$
{\cal Q}^{(1)}=-\int \frac{d^3P}{(2\pi)^3 2\omega_\mathbf{P}}\, \frac{(\omega_\mathbf{P}+P_4)^3}{P_4}\, |\phi^-(\mathbf{P})|^2
$$
and
$$
{\cal Q}^{(2)}=\int \frac{d^3P}{(2\pi)^3 2\omega_\mathbf{P}}\, \frac{P_4(\omega_\mathbf{P}+P_4)-1}{\omega_\mathbf{P}(P_4(\omega_\mathbf{P}+P_4)+\mathbf{P}^2)}\, |\phi^+(\mathbf{P})|^2\, ,
$$
where $\omega_\mathbf{P}=\sqrt{m^2+\mathbf{P}^2}$ and $\phi(\mathbf{P})\equiv \phi(k(\mathbf{P}))$. Note that ${\cal Q}$ does not depend on $x^0$, which was to be expected, of course, but it is nice to see explicitly.\\
Collecting terms and reintroducing the deformation parameter $\kappa$ we get\footnote{ As it can be easily checked the mass appears in this expression only through the combination $P_4=\sqrt{\kappa^2+m^2}$. Since $\kappa$ is expected to be of order of Planck mass, for the Standard Model mass scales we can safely neglect $m$. For this reason, and to make the formula below more transparent, we present it for massless case only. In this case $\mathbf{P}^2 =\omega^2_\mathbf{P}$. }
\begin{equation}\label{31}
    {\cal Q}=\int \frac{d^3P}{(2\pi)^3 2\omega_\mathbf{P}}\,\left\{\left(1+\frac{\omega_\mathbf{P}}{\kappa} +\frac{\omega^2_\mathbf{P}}{\kappa^2}\right)^{-1}\, |\phi^+(\mathbf{P})|^2 -\left(1+\frac{\omega_\mathbf{P}}{\kappa}\right)^3 \, |\phi^-(\mathbf{P})|^2 \right\}
\end{equation}
In the leading order of $1/\kappa$ expansion we get
\begin{equation}\label{32}
    {\cal Q}=\int \frac{d^3P}{(2\pi)^3 2\omega_\mathbf{P}}\,\left\{\left(1-\frac{\omega_\mathbf{P}}\kappa\right) |\phi^+(\mathbf{P})|^2 -\left(1+3\frac{\omega_\mathbf{P}}\kappa\right) |\phi^-(\mathbf{P})|^2 \right\}\, .
\end{equation}
Along the same lines one can easily calculate the expressions for the conserved translational charges (\ref{28}). To get ${\cal P}_A$ the only thing one has to do is to multiply the integrands in ${\cal Q}^{(1)}$ and ${\cal Q}^{(2)}$ above with $P_A$ and $S(P)_A$, respectively (cf.\ (\ref{14})). One finds (again we present the expressions for massless case)
\begin{equation}\label{33}
    {\cal P}_0=\int \frac{d^3P}{(2\pi)^3 2\omega_\mathbf{P}}\,\left\{\left(1+\frac{\omega_\mathbf{P}}{\kappa} +\frac{\omega^2_\mathbf{P}}{\kappa^2}\right)^{-1}\,  \left(\frac{\kappa \omega_\mathbf{P}}{\omega_\mathbf{P}+\kappa}\right) |\phi^+(\mathbf{P})|^2 +\left(1+\frac{\omega_\mathbf{P}}{\kappa}\right)^3 \, \omega_\mathbf{P}\, |\phi^-(\mathbf{P})|^2 \right\}
\end{equation}
\begin{equation}\label{34}
    {\cal P}_i=\int \frac{d^3P}{(2\pi)^3 2\omega_\mathbf{P}}\,\left\{\left(1+\frac{\omega_\mathbf{P}}{\kappa} +\frac{\omega^2_\mathbf{P}}{\kappa^2}\right)^{-1}\, \left(\frac{\kappa P_i}{\omega_\mathbf{P}+\kappa}\right)  |\phi^+(\mathbf{P})|^2 +\left(1+\frac{\omega_\mathbf{P}}{\kappa}\right)^3 \, P_i\, |\phi^-(\mathbf{P})|^2 \right\}
\end{equation}
As it turns out the fifth charge, ${\cal P}_4$ is actually proportional to ${\cal Q}$, which is clear from (\ref{28}) and (\ref{30}).

\section{Quantum field states: particles and antiparticles}
As anticipated in the previous Sections, the conserved charges we derived above are strictly related to the invariant symplectic product for classical solutions.  Such product, upon appropriate manipulation can be used to define a ``one-particle" Hilbert space for the quantum field and its full Fock space \cite{Arzano:2007ef}.  It turns out that the conserved $U(1)$ charge derived in the previous section corresponds to such symplectic product evaluated on the two independent fields $\hf^\dagger$ and $\hf$
  $$
 {\cal Q}=\omega(\hf^\dagger,\hf)= -\int_{\mathbb{R}^3}\left(\hat\Pi\hf +\hf^\dag\hat\Pi^\dag\right) \,.
  $$
  If we go back to the case of a real $\kappa$-scalar field it is easily seen that the symplectic product takes the form \cite{Arzano:2007gr}
\begin{equation}
\omega(\hf_1,\hf_2)=-\int_{\mathbb{R}^3}\left(\hat\Pi_1\hf_2 +\hf_1\hat\Pi_2\right)\, .
\end{equation}
The ``one-particle" Hilbert space for such real field can be constructed by restricting to positive energy solutions i.e. according to (\ref{27}) those solutions for which $\phi^{-}(\mathbf{P})\equiv 0$.  The (positive definite) inner product on such space will be given by
\begin{equation}
<\hf_1,\hf_2>=-i \omega(\overline{K^+\hf_1},K^+\hf_2)\, .
\end{equation}
where the operator $K^+$ projects real solutions on the positive energy subspace.  We can easily write the momentum space counterpart of such inner product
\begin{equation}
<\hf_1,\hf_2>=\int \frac{d^3P}{(2\pi)^3 2\omega_\mathbf{P}}\,\,\left(1+\frac{\omega_\mathbf{P}}{\kappa} +\frac{\omega^2_\mathbf{P}}{\kappa^2}\right)^{-1}
\overline{\phi_1^{+}(\mathbf{P})} \phi_2^+(\mathbf{P})\, ,
\end{equation}
notice that being the field real $ \overline{\phi^{+}(\mathbf{P})}$ and $\phi^{+}(\mathbf{P})$ are {\it not} independent.  For a complex field this is no longer the case and we will have two distinct inner products
\begin{equation}
<\hf_1,\hf_2>=\int \frac{d^3P}{(2\pi)^3 2\omega_\mathbf{P}}\,\,\left(1+\frac{\omega_\mathbf{P}}{\kappa} +\frac{\omega^2_\mathbf{P}}{\kappa^2}\right)^{-1}\,
\overline{\phi_1^{+}(\mathbf{P})} \phi_2^+(\mathbf{P})\, ,
\end{equation}
and
\begin{equation}
<{\hf_1}^\dagger,{\hf_2}^\dagger>=\int \frac{d^3P}{(2\pi)^3 2\omega_\mathbf{P}}\,\,\left(1+\frac{\omega_\mathbf{P}}{\kappa} +\frac{\omega^2_\mathbf{P}}{\kappa^2}\right)^{-1}\,
\phi_1^{-}(\mathbf{P}) \overline{\phi_2^{-}(\mathbf{P})}\, .
\end{equation}
Such inner products will define, respectively, the Hilbert space of ``one-particle states" $\mathcal{H}_p$ and ``one-antiparticle states" $\mathcal{H}_a$.\\  At the level of quantum operators the coefficients $\overline{\phi^{+}(\mathbf{P})}$ and $\phi^+(\mathbf{P})$ will be promoted respectively to creation and annihilation
operators for particle states, $a^{\dagger}(\mathbf{P})$ and $a(\mathbf{P})$, while
$\phi^{-}(\mathbf{P})$ and $\overline{\phi^-(\mathbf{P})}$ will give the antiparticle
counterparts $b^{\dagger}(\mathbf{P})$ and $b(\mathbf{P})$.\\
To conclude we take a look at the $U(1)$ charge carried by particle and antiparticle states.  The deformed quantum field theory counterpart of the conserved charge will
be given by the operators (for simplicity we restrict to the leading order in the $1/\kappa$ expansion)
\begin{equation}
    \hat{\cal Q}_p=\int \frac{d^3P}{(2\pi)^3
2\omega_\mathbf{P}}\,\left(1-\frac{\omega_\mathbf{P}}\kappa\right)
a^{\dagger}(\mathbf{P}) a(\mathbf{P})\, .
\end{equation}
and
\begin{equation}
    \hat{\cal Q}_a=\int \frac{d^3P}{(2\pi)^3
2\omega_\mathbf{P}}\,\left(1+3\frac{\omega_\mathbf{P}}\kappa\right)
b^{\dagger}(\mathbf{P}) b(\mathbf{P})\,
\end{equation}
The expectation values of these operators in one-(anti)particle states will correspond to their ``electric charges". We notice that while particle and antiparticle states carry the same mass (as discussed earlier in Section II), at first sight it would seem instead that their charges are not only different, but also energy dependent. This conclusion is, however, not justified at this stage. The reason being that the expectation value of $\hat{\cal Q}_p$, say, in a state of momentum $\mathbf{K}$ is given by
\begin{equation}
 \left<\hat{\cal Q}_p \right> =  \left<\mathbf{K}\left|\hat{\cal Q}_p\right|\mathbf{K}\right> = \left<0\left|a(\mathbf{K})\,\hat{\cal Q}_p\,a^\dagger(\mathbf{K})\right|0\right>
\end{equation}
In order to calculate this we would need to know the exact form of the commutators of the annihilation/creation operators, which will be deformed as well as a result of the nontrivial coproduct for the mode labels $\mathbf{K}$ (see \cite{Arzano:2008bt} for an extended discussion). We will return to this issue in a forthcoming work.

\section{Conclusions}
The aim of this paper was to construct an example of a field theory that does not suffer from the global Lorentz symmetry breaking problem noticed in \cite{Freidel:2007hk}, \cite{Freidel:2007yu}. We achieve this by labeling plane waves corresponding to the  negative energy states by antipodes of momenta, not by the ``minus'' momenta. As we has shown this cures the problem. The price we have to pay was that the Lorentz group action on momenta labeling positive and negative energy states is not the same now.\\
As an example we studied a free complex scalar field starting from its Lorentz invariant action.  Its equations of motion correspond to the standard dispersion relation and the mass for positive and negative energy modes turns out to be the same.  This led, in the quantum framework, to the equivalence for the masses of particles and antiparticles.\\
We then derived the expression for the deformed conserved translational and $U(1)$ charges. The interest in the latter charge is twofold. First, in the quantum theory, the corresponding operator measures the electric charges of one particle states, and therefore it is required in the discussion of the fate of $C$, and $CPT$ symmetries in a $\kappa$-deformed setting.  Second, as discussed in Section V such charge is directly related to the inner product which defines the quantum fields Hilbert space and thus is essential in constructing the corresponding deformed two-point function. We will address these issues in a forthcoming work.

\begin{acknowledgments}
JKG would like to thank Perimeter Institute for hospitality while this work has been completed.  Research at Perimeter Institute is supported by the Government of Canada through Industry Canada and by the Province of Ontario through the Ministry of Research \& Innovation.  JKG is supported in part by research projects N202 081 32/1844 and NN202318534 and Polish Ministry of Science and Higher Education grant 182/N-QGG/2008/0.
\end{acknowledgments}

\end{document}